\documentclass[11pt]{aastex}
\usepackage{epsfig}
\newcommand{\ms}{$M_{\odot}$}

\shorttitle{IR Photometry of AGB Stars}
\shortauthors{Busso et al.}
\begin{document}
\title{Mid Infrared Photometry of Mass-Losing AGB Stars.}
\author{M Busso \altaffilmark{1}}
\affil{Department of Physics, University of Perugia, via Pascoli,
Perugia, Italy, 06123; busso@fisica.unipg.it}

\author{R. Guandalini\altaffilmark{1}}
\affil{Department of Physics, University of Perugia, via Pascoli,
Perugia, Italy, 06123; guandalini@fisica.unipg.it}

\author{P. Persi \altaffilmark{2}}
\affil{INAF, Istituto di Astrofisica Spaziale e Fisica Cosmica,
via Fosso del Cavaliere, 00100 Roma, Italy; persi@iasf-rm.inaf.it}

\author{L. Corcione\altaffilmark{3}}
\affil{INAF, Osservatorio Astronomico di Torino, via Osservatorio
20, 10025 Pino Torinese, Italy; corcione@oa-torino.inaf.it}

\and

\author{M. Ferrari-Toniolo\altaffilmark{2}} \affil{INAF, Istituto di
Astrofisica Spaziale e Fisica Cosmica, via Fosso del Cavaliere,
00100 Roma, Italy; ferrari@iasf-rm.inaf.it}
%
%

\begin{abstract}
We present ground-based mid-infrared imaging for 27 M-, S- and
C-type Asymptotic Giant Branch (AGB) stars. The data are compared
with those of the database available thanks to the IRAS, ISO, MSX
and 2MASS catalogues. Our goal is to establish relations between
the IR colors, the effective temperature $T_{eff}$, the luminosity
$L$ and the mass loss rate $\dot M$, for improving the
effectiveness of AGB modelling. Bolometric (absolute) magnitudes
are obtained through distance compilations, and by applying
previously-derived bolometric corrections; the variability is also
studied, using data accumulated since the IRAS epoch. The main
results are: i) Values of $L$ and $\dot M$ for C stars fit
relations previously established by us, with Miras being on
average more evolved and mass losing than Semiregulars. ii)
Moderate IR excesses (as compared to evolutionary tracks) are
found for S and M stars in our sample: they are confirmed to
originate from the dusty circumstellar environment. iii) A larger
reddening characterizes C-rich Miras and post-AGBs. In this case,
part of the excess is due to AGB models overestimating $T_{eff}$
for C-stars, as a consequence of the lack of suitable molecular
opacities. This has a large effect on the colors of C-rich sources
and sometimes disentangling the photospheric and circumstellar
contributions is difficult; better model atmospheres should be
used in stellar evolutionary codes for C stars. iv) The presence
of a long-term variability at mid-IR wavelengths seems to be
limited to sources with maximum emission in the 8 -- 20 $\mu$m
region, usually Mira variables (1/3 of our sample). Most
Semiregular and post-AGB stars studied here remained remarkably
constant in mid-IR over the last twenty years.
\end{abstract}

\section{Introduction}

Stars of low and intermediate mass (all those below M = 7-8 \ms)
terminate their evolution through the so-called Asymptotic Giant
Branch (AGB) phase \cite{bus99,bus07}, in which they lose mass
efficiently thanks to stellar winds powered by radiation pressure
on dust grains \citep{habing}. After this stage, they generate
planetary nebulae and start a blue-ward path, which ultimately
gives birth to a white dwarf \citep[see also][]{herwi05}.

While moving along this path, AGB stars replenish the Interstellar
Medium with about 70\% of all the matter returned after stellar
evolution \citep{sedlmayr94}; this is done through the formation
of extended circumstellar envelopes \citep{winters02}. Cool or
cold dust, with its radiative emission in the infrared (IR),
normally dominates the energy distribution of these sources, so
that until recently the bolometric (apparent) magnitude of the
most evolved AGB stars was difficult to derive, due to
insufficient photometric coverage of the mid-far IR range of the
electromagnetic spectrum. Traditionally, circumstellar envelopes
were studied using fluxes provided by the IRAS satellite
\citep[see][]{jura86,jura89}, which had however a poor spatial
resolution (hence some risk of contamination, especially at the
longest wavelengths). Similarly, absolute magnitudes were very
uncertain due to the difficulties in measuring the distance of
these single, strongly variable stars.

Only recently, the availability of a large IR database from
space-borne telescopes like ISO, IRTS, MSX and the increased
amount of ground-based mid-IR observations have substantially
modified the situation. The quantitative studies of luminosities,
colors and mass loss in the last evolutionary stages of moderately
massive stars have consequently grown in number and quality
\citep{whitelock,Feast06}. In parallel, Hipparcos distances for
AGB stars have been in part upgraded \citep{berg05}. The results
may still include inconsistencies (Feast, private communication)
but a full revision of the Hipparcos catalogue is still waited for
\citep{val05,valfan05}. Recent works have therefore largely
exploited Period-Luminosity relations to infer the distances
\citep{Menzies06,feast07}. Contemporarily, the first results of
Spitzer's surveys are becoming available \citep{zijl06}.

In the above modified scenario we started a reanalysis of galactic
AGB stars at IR wavelengths, aimed at improving the determination
of their energy distribution, mass loss and absolute magnitudes,
making use of recent IR photometric data and of reliable distance
estimates. With such works one can now finally compare homogeneous
samples of mass losing stars in the Galaxy with similar data sets
in Magellanic Clouds \citep{vanloon01, vanloon05}, in order to
study the dependence on metallicity of general stellar properties.

In a previous paper \citep[][ hereafter Paper I] {guan06} we
considered C stars as observed by the ISO-SWS and by the MSX
experiments. Here we extend our work by analyzing a sample
 of 27 AGB stars of classes M, S and C (listed in Table 1) for which we obtained
ground-based infrared imaging in the 10 $\mu$m window. Most of the
stars we observed were also the object of measurements by the
above quoted space-borne IR telescopes, so that we can now compare
and integrate results from different experiments and different
epochs, as a check of the quality of the available infrared
database and of the source variability.

This paper is organized as follows: in section 2 we briefly
discuss our IR camera, used for making the ground based
observations. In section 3 we present the photometric data
obtained through it at wavelengths longer than 8 $\mu$m,
integrated by the near-IR archive observations of 2MASS
\citep{cutri} and by the available estimates for mass loss rates,
distances and a few other relevant parameters. In section 4 we
then use the database thus constructed to derive bolometric
magnitudes (both apparent and absolute), colors, and correlations
of photometric properties with mass loss rates. Then section 5
addresses the long-term variability issue, by going back to the
IRAS catalogues in order to compare the available data over a time
interval of about 20 years. Finally, in section 6 we outline some
general conclusions, underlining the main remaining problems that
remain to be solved for allowing a satisfactory match between
photometric observations and stellar modelling.

\section{The Mid-IR Camera TIRCAM2 and its Calibration}

Our sample stars were observed between 2001 and 2004 at the
Italian Infrared telescope (TIRGO), located at 3200 meters over
the sea level, on top of the Gornergrat, in Switzerland. The data
were collected using our mid infrared camera TIRCAM2 (Tirgo
Infrared Camera - version 2), an upgrade of a previously available
instrument \citep{persi94}.

TIRCAM2 uses a Rockwell (then Boeing) high-flux Si:As BIB 128
$\times$ 128 array (HF-21) and is equipped with five narrow-band
filters (10\% bandwidth) between 8 and 13 $\mu$m, with the N
broad-band filter, and with a circular variable filter (CVF)
having a spectral resolution of 3\% in the 8-14 $\mu$m range. The
optics of the camera produce a plate scale at TIRGO of
0.77\arcsec/pix. The array, the optical system and the filters are
assembled in a liquid-He-cooled dewar (HD-3[8]) of the Infrared
Labs. Inc. (Tucson, Arizona). The readout electronics of the
array, fully developed by our home Institutes, and the other
general characteristics of the instrument have been previously
described elsewhere \citep{persi02,corc03}. The absolute fluxes of
some infrared standard stars in our photometric system are shown
in Table 2. They have been derived from the spectral energy
distributions (SEDs) published by Cohen and coworkers \citep[][and
references therein]{cohen}, after a convolution with the spectral
response of our filters \citep{persi02}.

The narrow-band images at 8.8, 9.8, 11.7 and 12.5 $\mu$m of our
sample of AGB stars were taken in the standard chopping \& nodding
technique, in order to remove the sky and telescope emission
background. Images of standard stars were obtained during the
nights at similar air mass as our targets, for flux calibration.
They were also used to estimate the point-spread function (PSF) at
each wavelength. The mean PSF (FWHM) obtained during our observing
runs was $\sim$ 3.2$\arcsec$ in diameter at 8.8 $\mu$m and $\sim$
3.8$\arcsec$ at 12.5 $\mu$m. All the observed AGB stars appear as
point-like sources at this spatial resolution. The on-source
integration times were from 720sec for the faintest sources, down
to 120sec for the brightest ones, in all the filters used. The
derived detection limit is approximately 0.7 Jy (1 $\sigma$) in
300sec of integration time. The photometric measurements were
extracted from the images of AGB stars through the DAOPHOT tool,
within the IRAF data reduction package \citep{ste87}, using an
aperture of 6\arcsec.

\section{Infrared Colors and Mass Loss of AGB Sources}

The flux densities of the sources observed, as measured with the
technique mentioned in the previous section, are given in Table 3,
together with their 1$\sigma$ statistical errors. We collected, as
a comparison, near infrared fluxes for our sources, as published
in the 2MASS catalogue, together with observations by the MSX
satellite (when available) for wavelengths longer than those of
our filters. These data are presented in Table 4. Finally,
distances, mass loss rates and wind velocities, as deduced by the
literature, are presented in Table 5, together with the relevant
references.

By calibrating the data of Table 3 with the zero-magnitude fluxes
deduced by standard stars in Table 2, we obtain the infrared
colors of the sources. Examples of color-color diagrams are shown
in Figure~\ref{fig:one} and in Figure~\ref{fig:two}. Here TIRCAM2
sources are represented by big symbols: open-starred ones refer to
M-type (oxygen-rich) stars, plain open ones to S-type stars.
Filled symbols indicate C-rich sources. Small dots refer to the
sample of C stars we discussed in Paper I, with colors deduced
from ISO and MSX observations, as a comparison (see Figure 6 in
Paper I). We recall that the ISO data were taken from the last
on-line database available, where SWS measurements are fully
calibrated \citep{Leech}. From them we derived photometric
estimates in the TIRCAM2 photometric system by convolving SWS
spectra with the response curves of our filters (see Paper I).

The bulk of our data is generally distributed along the line of
black body emission in Figure \ref{fig:one}, while a larger
dispersion is present in Figure \ref{fig:two}.

The spread in Figure \ref{fig:two} derives from the fact that, for
sources that are presently on the AGB (Semiregulars or Miras),
color indexes computed in the $8 - 14 \mu$m region are affected by
the crowding of variable emission (or absorption) features
characteristic of these wavelengths (e.g. at $11.2 - 11.7 \mu$m
for C-stars, from PAH and SiC; and at 9.8 $\mu$m for O-rich
sources, from silicates). When colors outside the $8-14 \mu$m band
are used (Figure \ref{fig:one}) excess emission at 21 $\mu$m is
seen for a number of post-AGB, C-rich TIRCAM2 sources (as also
shown by ISO and MSX data, see Paper I). Emission at this
wavelength was known to be a distinctive property of a few
post-AGB C stars and C-rich pre-planetary nebulae \citep{kwok02,
zha06}. However, so far the number of known sources with this
feature was rather limited: in our data (both here and in Paper I)
the 21 $\mu$m emission appears as a general property of post-AGB
sources. The above simple considerations confirm the fact that
mid-IR colors are effective tools for classifying AGB sources.
When one deals with infrared stars, mid-IR colors identify C-rich
AGBs through their color excess at 11.7$\mu$m, and C-rich
post-AGBs through their 21 $\mu$m excess.

Figure \ref{fig:three} then shows the relations linking the IR
colors to mass loss. Again, the best measurements selected for C
stars in Paper I are included for comparison. It is clear that our
new C-star data follow the trends already established, as
expected. It is however also evident that the (few) O-rich sources
of the sample behave differently, displaying lower mass-loss rates
for a given value of the color, especially when this last makes
use of fluxes in the 10$\mu$m region (panel "a" in Figure
\ref{fig:three}). Should this hint be confirmed, this would mean
that the relation between dust opacities and mass loss rates for M
and S stars has to be different than for C-stars. We are now
verifying this in a dedicated work on a much larger database
(Guandalini \& Busso in preparation).

\section{Color-magnitude Diagrams and Effective Temperatures}

By applying to our database the bolometric corrections discussed
in Paper I, we can easily derive apparent bolometric magnitudes of
the C-rich sources studied. For O-rich S and M stars we applied
instead the bolometric correction \cite{guan07}:
$$
M_{bol} - M[8.8] = a x^3 + b x^2 + c x + d \eqno(1)
$$
where $x$ is the K-[8.8] color, and the coefficients are
a=$-$0.0211, b=0.0812, c=1.0658, d=2.3026 (with a correlation $r^2
= 0.989$). Correcting for the distance (see Table 5) yields the
absolute bolometric magnitude, through which the absolute
color-magnitude diagram can be plotted. Examples are presented in
Figures~\ref{fig:four} and ~\ref{fig:five}.

In Figure~\ref{fig:four} the dashed area represents the zone
occupied by AGB photospheres according to current stellar models
\citep{stra97}. As the Figure shows, essentially all the stars of
our sample, much like those of Paper I, are reddened with respect
to the models, by amounts that are widely disperse. Especially for
the reddest stars, the displacements are dominated by the known
presence of dust in the circumstellar envelope. In this respect,
an additional source of concern is however introduced by the poor
knowledge of molecular opacities for C-rich atmospheres, and by
the fact that they are usually neglected in stellar codes
\citep{marigo}. Due to this, it is known that the model effective
temperatures of very cool, C-rich AGB models, in the region of
thermal pulses, are unreliable and generally largely
overestimated. For strongly variable, dynamically perturbed
atmospheres the same physical meaning of $T_{eff}$ becomes
doubtful \citep{utt07}.

Uncertainties in model $T_{eff}$ values have very different
consequences on the colors of M-type and C-type stars. As an
example, let's consider, for AGB stars of the two classes, the
effects of a change by 0.1 dex in $\log ~T_{eff}$. For M stars
this corresponds e.g. to moving from type M3 to type M8: the (J-K)
color difference in Figure~\ref{fig:four} is less than 0.2mag. On
the contrary, using for C stars the classifications and the color
calibrations by Bergeat et al. (2001), we see that the same shift,
corresponding e.g. to moving from the class CV6 to the class CV7
\citep{bergeat02}, may imply a color difference (J-K) $\simeq$ 1
\citep[see Figure 4 in][]{bergeat01}. Similar considerations would
hold for the [K-12.5] color in Figure~\ref{fig:five}. We conclude
that, while the colors of O-rich AGB stars are not significantly
affected by uncertainties in the atmospheric models, for C-stars
the situation is worse. At least the less reddened stars in
Figure~\ref{fig:four} might show displacements from the model
areas partly due to the real presence of dust, but partly also
induced by errors in model $T_{eff}$ values, hence in atmospheric
opacities.

A special word of caution must be added for the positions of
post-AGB objects in the two plots discussed here. Sometimes they
have a complex SEDs, related to the fact that the central star
begins to be detached from the circumstellar shell and shines at a
relatively high temperature, as typical of a yellow supergiant. In
such cases the integral of the IR flux (and in general the
procedure described in Paper I) is not sufficient to determine the
bolometric magnitude properly, because the optical contribution is
not negligible \citep[see e.g.][ Fig. 1]{kwok99}. For post-AGB
objects we have therefore only lower limits to the bolometric
magnitudes, which approximate the real values with an error that
is different for different cases.

In order to illustrate the cautious remarks made above for the
relations between the color excess and the presence of dust, we
selected in our sample of TIRCAM2 sources, and in the "best set"
of sources in Paper I (those with astrometric distances and very
detailed SEDs up to beyond 40 $\mu$m) a group of stars for which
independent estimates of $T_{eff}$ exist
\citep{olofsson93a,olofsson93b,bergeat01} or can be inferred from
published color calibrations \citep{bessell}. For these objects we
plot the absolute HR diagram ($M_{bol}$ versus $T_{eff}$) in
Figure \ref{fig:six}. In the Figure, the tracks on the left refer
to AGB models for various masses and metallicities \citep{stran3}.
It is evident that, while the observed magnitudes fall in the
range expected by these calculations (which include minimal or no
overshooting), a large number of sources, and in particular the
C-rich ones, have temperatures much lower than those predicted:
these last, lacking a treatment for C-rich molecular opacities,
can be in error by up to 30\%.

The effects of absorption features from molecules like CO, CN and
C$_2$ was shown to be extremely sensitive to the effective
temperature and of large consequence for the near infrared colors
by an hydrostatic analysis of a few C stars \citep{loidl}. The
influence of molecular opacities for the changing composition of
AGB envelopes gradually enriched by the TDU process, and in
particular for C-rich sources, was also addressed by Marigo
(2006). She showed how a large decrease in $T_{eff}$ can be
induced by the inclusion of opacities from molecules made of CNO
and hydrogen. In models of about 2 $M_{\odot}$ a shift in
$Log~T_{eff}$ of almost 0.1 dex can be obtained (see Figure 6 in
that paper). Just as an example, by reducing of this amount the
temperatures of the model tracks in Figure \ref{fig:six} (in order
to simulate the effect of molecular opacities) we get the curves
at the right side of the plot (those with crosses superimposed).
It is evident that a proper atmospheric model might be in many
cases sufficient to yield the values of $T_{eff}$ derived from
observations. We warn that the new tracks must be seen only as a
rough example, plotted for illustration purposes. In fact, Marigo
(2006) computed a synthetic AGB evolution (while the tracks in
Figure \ref{fig:six} were derived from complete AGB models), and
her assumptions for dredge-up were rather different than those in
the models we used \citep{stran3}. What one would really need here
is a set of complete AGB models, including a proper treatment of
surface opacities for the changing envelope composition during the
TDU process, something that unfortunately does not exist yet.

The above discussion says that, for C-stars, disentangling the
atmospheric opacity effects from the color excess due to dust will
become quantitatively meaningful only with large samples of good
mid-IR observations, determining the properties of dust for
sources where $T_{eff}$ has been independently determined (e.g.
from spectra). This in its turn would help stellar modelers to
construct more detailed opacity tables, calibrated on
observations. This is a relevant target to be pursued. In fact,
the compilation of accurate infrared catalogues was started by
several groups years ago \citep[see for
example][]{vanloon99,cioni01,lebertre01}. It has continued to be
an essential tool in recent years
\citep{bergeat02,lebertre03,cioni03} and remains important now
\citep{berg05,lebertre05,whitelock}. It will also be one of the
key projects of the Antarctic telescope IRAIT (International
Robotic Antarctic Infrared Telescope) that we recently developed
and that will be operative at the Italo-French base of Dome C
starting from the 2007-2008 Antarctic campaign \citep[see
e.g.][]{tosti, bus06}.

\section{Infrared Variability}

Figures \ref{fig:seven} and \ref{fig:eight} show the available
information on the IR SEDs for two groups of sources in the list
of Table 1. The plots include data from the IRAS-PSC (Point Source
Catalogue), IRAS-LRS (Low Resolution Spectra), ISO and MSX,
together with our TIRCAM2 measurements. Figure \ref{fig:seven}
shows distributions that, despite their different appearance,
share the property of being non-variable over a time interval of
almost 20 years. The figure contains rather heterogeneous
measurements: photometric data (from the IRAS Point Source
Catalogue, from MSX and from our ground-based observations) are
compared with spectroscopic information from ISO-SWS and IRAS-LRS.
For our purposes this is sufficient: as verified in Paper I, the
proper convolution of ISO-SWS and IRAS-LRS spectra with the
response of our filters, yielding a homogeneous photometric
database, would not affect the flux levels by more than 5\%, which
is well inside the internal uncertainty of each set of data used.
A constancy of the IR energy distribution is the more common
behavior displayed by our sources, being shared by exactly 2/3 of
the AGB stars in our sample (18 out of 27). This property
characterizes stars that are very different from one another. They
include sources with minimal IR excess (usually Semiregulars), in
which the SED is peaked in near-IR; but they also include evolved
(post-AGB) objects, in which the maximum emission is at very long
wavelengths (from 20 to more than 40 $\mu$m), due the dominant
effects of cold, distant dust.

In contrast, Figure~\ref{fig:eight} shows the behavior of
intermediate objects, usually Mira variables, in which the
emission peaks near 10 $\mu$m (and sometimes is rather flat up to
about 20$\mu$m), efficiently powering the typical features there
present for O-rich and C-rich dust. Nine sources share this
behavior (RAFGL sources n. 190, 809, 865 and 954; IRC+60144; RU
Vir; CIT 6; CW Leo; IRC+40156). It seems, therefore, that
long-term mid-IR variability is not a common property of AGB
stars, but is restricted to a special class of sources, in the
special evolutionary stage when most of the flux is re-radiated by
circumstellar layers of one to a few hundred K. Here stellar
pulsation (which is of large amplitude in the optical bands)
should effectively transfer energy to dust shells, which are quite
opaque down to the mentioned temperatures. One can guess that this
is the phase where the coupling between the photosphere and the
circumstellar envelope is most efficient: for bluer sources, not
enough dust is created, probably by a mass loss rate which is
still not strong enough. For redder objects, circumstellar dust
becomes cold and distant, probably detaching itself from the
central star.

One might a priori argue that the Semiregular sources might look
stable only because of their small IR fluxes: when the flux is
very low any variability might be more difficult to disentangle
from the background noise. In order to make our suggestions more
secure we looked for sources (both variable and non-variable) for
which the ISO-SWS instrument offered repeated observations. We
found a few interesting cases. Examples are presented in
Figure~\ref{fig:nine} for the C-rich semiregular star S Scl and
for the C-rich Mira variable V Cyg. Even looking, in the figure,
only at homogeneous data sets (those from ISO, covering about 2
years) our previous suggestions seem to be confirmed. It is indeed
clear that the Mira variable does vary as we mentioned, and it can
also be understood how the non-variability of S Scl in
mid-infrared in the time interval covered by the ISO data (at a
level better than 10\%) does not descend from an insufficient
precision, since the minimum flux is of the order of tens of Jy.

The idea that the variable sources are those for which radiation
pressure on dust grains powers mass loss suggests itself, though
we can look at this only as at a tentative interpretation, because
the source statistics and also the number of available points per
source are rather limited. This hypothesis deserves now to be
verified through modelling of the radiation transfer in the
dust-condensing envelope, also to determine whether IR variability
corresponds to a specific evolutionary stage, or can be
encountered repeatedly (the same dilemma presented by optical Mira
variability).

\section{Conclusions}

The data presented in this note confirm the usefulness of mid-IR
colors taken from ground based telescopes, in describing the
properties of mass-losing AGB stars. Our photometric study yields
results compatible with those of Paper I, and begins to extend the
analysis to a number of O-rich (S and M) sources. These will now
be the object of dedicated papers, based on a large space-borne
observational database similar to that of Paper I.

In general we showed how, for most semiregular variables and
post-AGB stars, the IR fluxes longward of a few microns are
insensitive to the variations in the stellar photosphere, and
remain essentially constant in time over rather long time
intervals (tens of years). This however does not apply to Mira
variables, and in general to AGB sources having their maximum
emission near 10 $\mu$m. Such objects show remarkable variability
in the emission/absorption features (with changes from emission to
absorption and back) and/or in the global flux, showing that warm
dust quickly reacts to the large-amplitude surface pulsations,
with large changes in concentration and temperature. This might
suggest that dust-driven winds are mainly associated to the Mira
stage, a hypothesis to be further verified.

Our results also point out how urgent it is to match stellar
evolutionary codes with reliable model atmospheres, especially for
carbon stars, including molecular opacities. This would allow one
to predict reliably the effective temperature of C-rich
atmospheres and hence to give a quantitative meaning to the IR
color excess in terms of dust emission, without remaining
uncertainties from poorly understood photospheric opacities.

{\bf Acknowledgements}.

M.B. and R.G. acknowledge support by MIUR (contract
PRIN2004-025729) and by PNRA (within the IRAIT project). R.G.
acknowledges the University of Perugia for a post-doc fellowship.
TIRCAM2 was operated by IASF-CNR and by the Observatory of Torino
(both are now part of INAF). A special thank goes to A. Ferrari,
for supporting the TIRCAM2 project during his term as a director
of the Torino Observatory.

This research made use of the SIMBAD database, of the VizieR
service (CDS, Strasbourg, France) and of the Astrophysics Data
System of NASA. In particular, archived data from the experiments
MSX, ISO-SWS and 2MASS were used. [The processing of the science
data from the Midcourse Space Experiment (MSX) was funded by the
US Ballistic Missile Defense Organization with additional support
from the NASA Office of Space Science. The Infrared Space
Observatory (ISO) was an ESA project with instruments funded by
ESA Member States (especially the PI countries: France, Germany,
the Netherlands and the United Kingdom) and with the participation
of ISAS and NASA. The Two-Micron All-Sky Survey (2MASS) was a
joint project of the University of Massachusetts and of the
Infrared Processing and Analysis Center (IPAC) at the California
Institute of Technology; it was funded by NASA and by the NSF
(USA)].

\newpage
\centerline{\bf   FIGURE CAPTIONS}

\figcaption{An example of a color-color diagram for the observed
stars. Big symbols are from TIRCAM2, small ones from Paper I. See
text for comments.\label{fig:one}}

\figcaption{Another example of a color-color diagram for the
observed stars. Again, big symbols are from TIRCAM2, small ones
from Paper I. See text for explanations.\label{fig:two}}

\figcaption{Mass loss rates as a function of the infrared colors
for the sample of stars in this paper (large dots) and for the
best-measured sources of Paper I (small dots). As usual, open
symbols refer to S stars, filled symbols to C
stars.\label{fig:three}}

\figcaption{The color-magnitude diagram for our sources, using the
J-K color as a temperature indicator (see text for explanations).
Small symbols are from Paper I. The "C-star limit" line refers to
2 $M_{\odot}$ models of solar metallicity.\label{fig:four}}

\figcaption{Same as above, but using the K-[12.5] color in
abscissa. This color samples dust of different temperatures (and
also, in K, the residual photospheric flux); the horizontal
extension of the data is a measure of the infrared excess due to
the dusty envelope.\label{fig:five}}

\figcaption{The HR diagram of the few TIRCAM2 C-rich (full large
dots) or S-type (open large symbols) stars for which we could find
in the literature independent estimates of $T_{eff}$. For S-stars
$T_{eff}$ is an extrapolation \citep[from the relations
by][]{bessell}. Small symbols refer to those C-rich stars from
Paper I having astrometric distances and values of $T_{eff}$. The
curves at the right are from the AGB models at various masses and
metallicities \citep{stran3}. It is evident that, on average,
C-rich sources show large temperature discrepancies compared to
models sequences. The curves on the right side are shifted by
0.1dex in $Log ~T_{eff}$, in order to mimic molecular opacity
effects \citep{marigo1}.\label{fig:six}}

\figcaption{The Spectral Energy Distributions of a few
non-variable sources, as available from our TIRCAM2 observations
and from the IRAS PSC, IRAS LRS, ISO-SWS, MSX, and 2MASS
catalogues. SEDs with maximum emission in the near-infrared, as
well as those with maximum emission longward of 20$\mu$m all show
a constant flux in mid-IR.\label{fig:seven}}

\figcaption{Spectral Energy Distributions of sources that show
significant variability over the time elapsed from the IRAS to the
TIRCAM2 observations. Again, data obtained with TIRCAM2 and data
available from the IRAS PSC, IRAS LRS, ISO-SWS, MSX, and 2MASS
catalogues are included. Only sources for which the emission is
maximum in the range $8-20 \mu$m do appear to be variable,
independently on their composition (C-rich or
O-rich).\label{fig:eight}}

\figcaption{ISO-SWS Spectral Energy Distributions of two Carbon
Stars: a Mira variable (top panel) and a Semiregular (bottom
panel); they were observed repeatedly by ISO during about 2 years.
On this rather short time scale, but with homogeneous data, our
suggestions about mid-IR variability of Miras and flux constancy
of Semiregulars seem to be confirmed.\label{fig:nine}}
%
\clearpage

\begin{figure}
\begin{center}\includegraphics[width=0.80\columnwidth,angle=-90,clip]{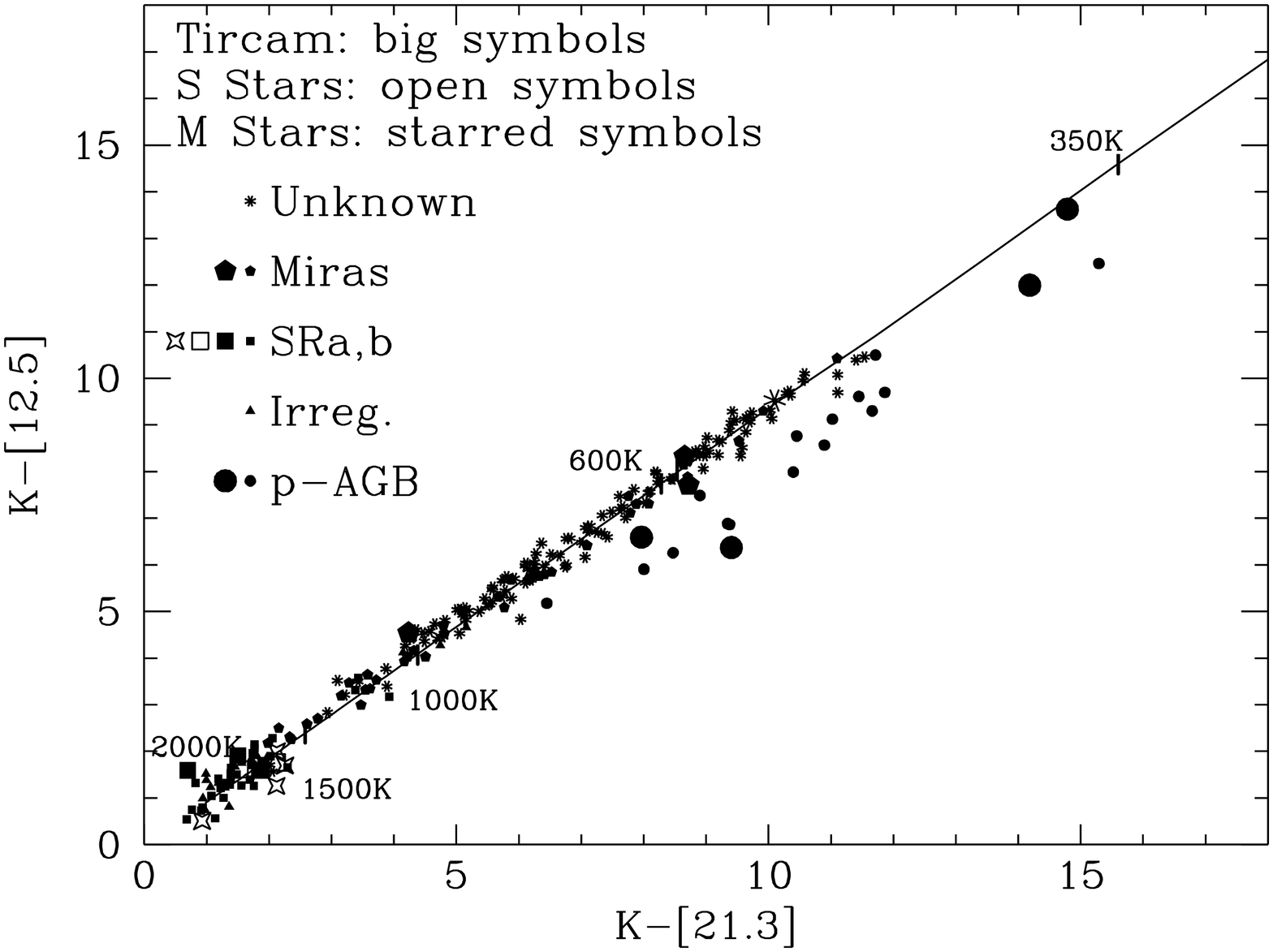}

\end{center}

\end{figure}

\begin{figure}
\begin{center}\includegraphics[width=0.80\columnwidth,angle=-90,clip]{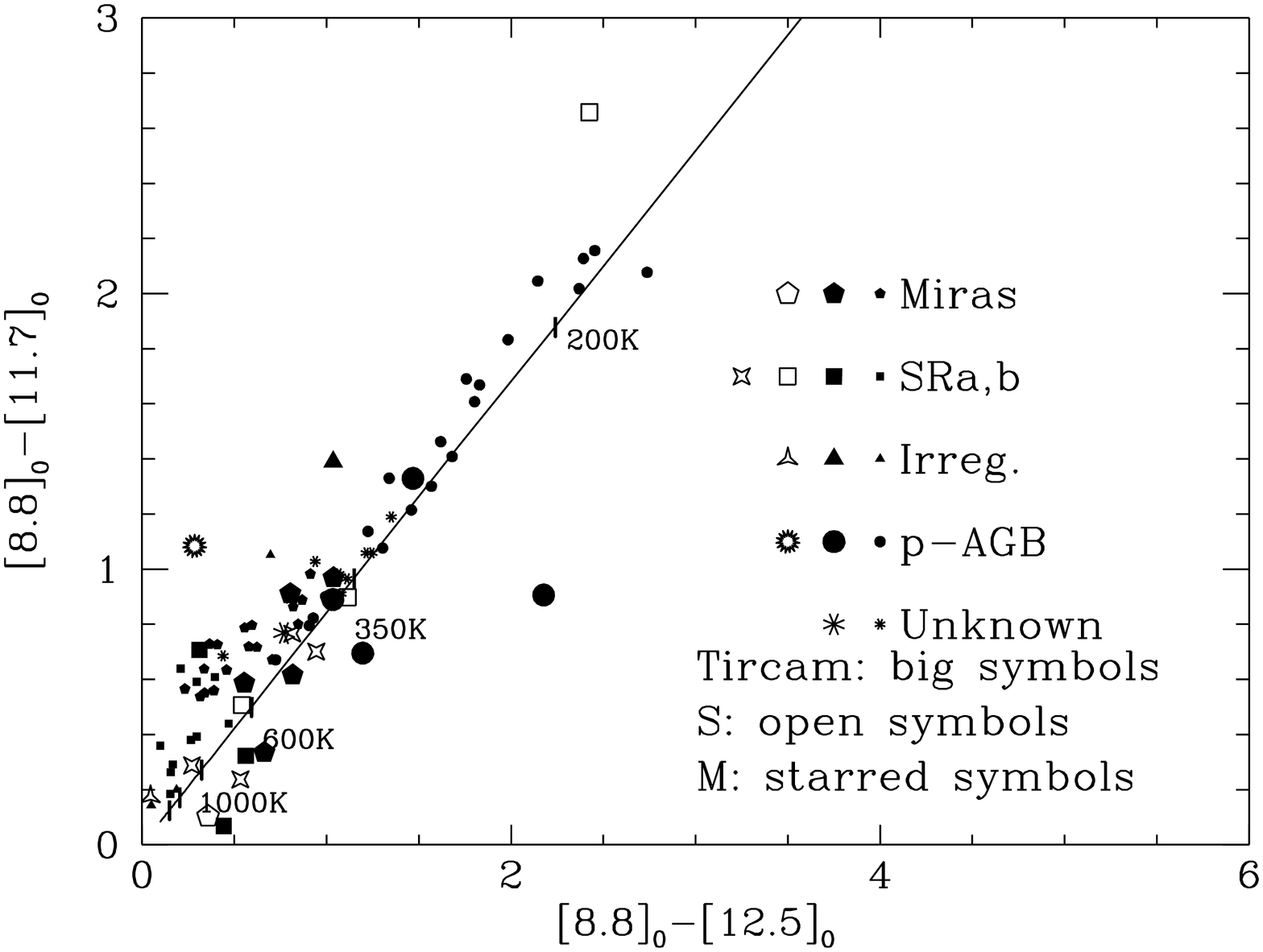}
\end{center}

\end{figure}

\begin{figure}
\begin{center}\includegraphics[width=0.80\columnwidth,clip]{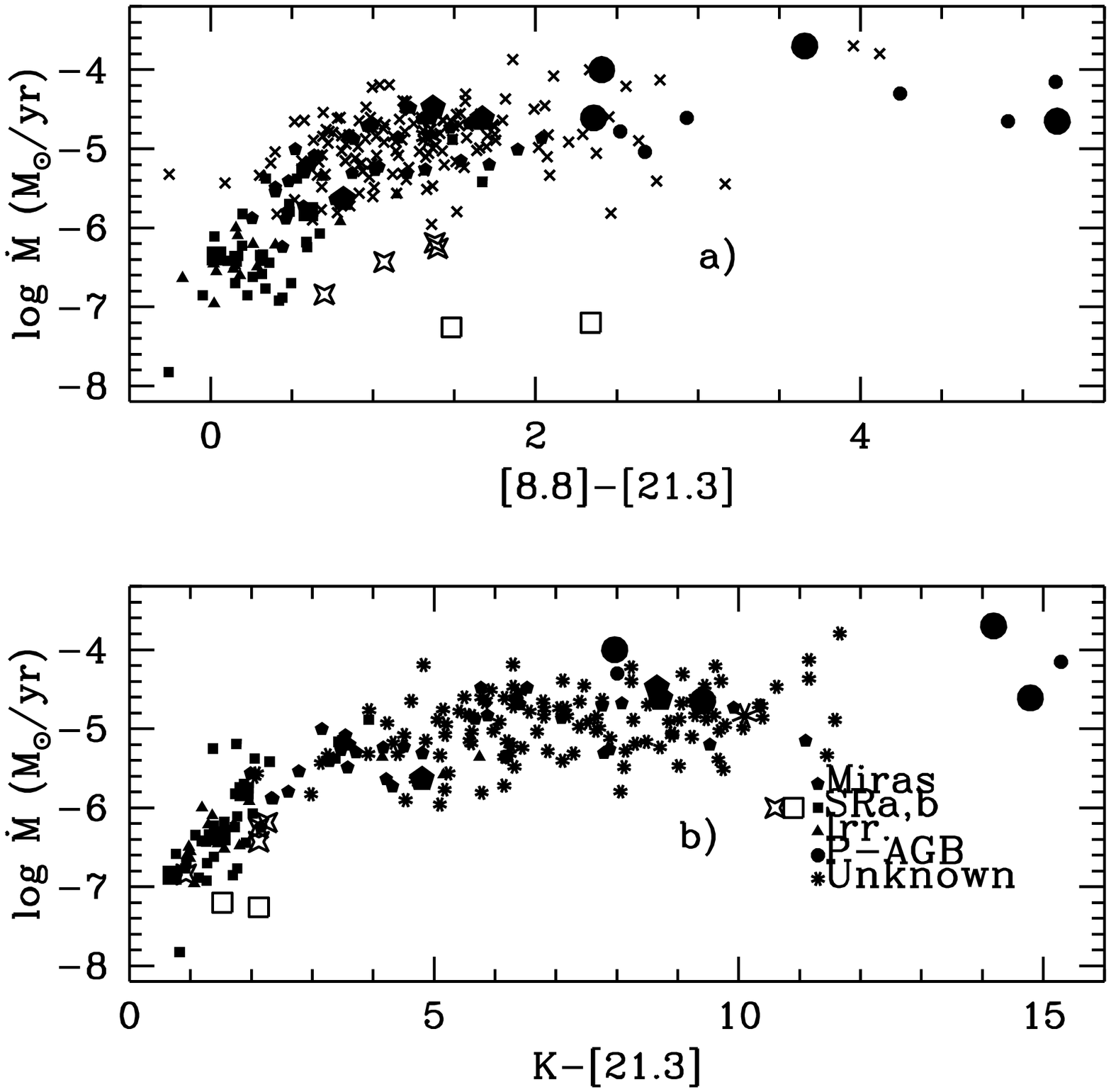}
\end{center}

\end{figure}

\begin{figure}
\begin{center}\includegraphics[width=0.80\columnwidth,angle=-90,clip]{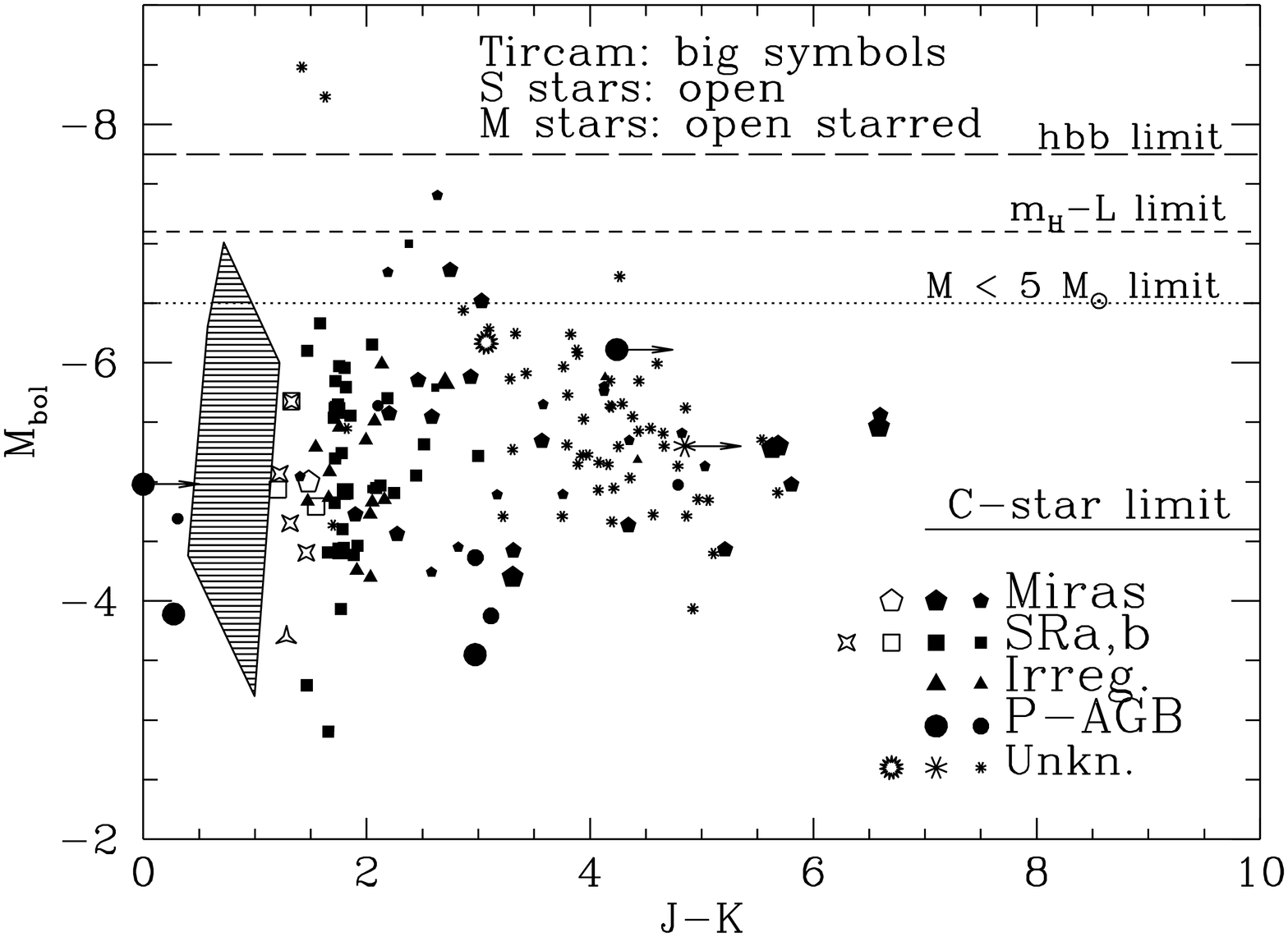}
\end{center}

\end{figure}

\begin{figure}
\begin{center}\includegraphics[width=0.80\columnwidth,angle=-90,clip]{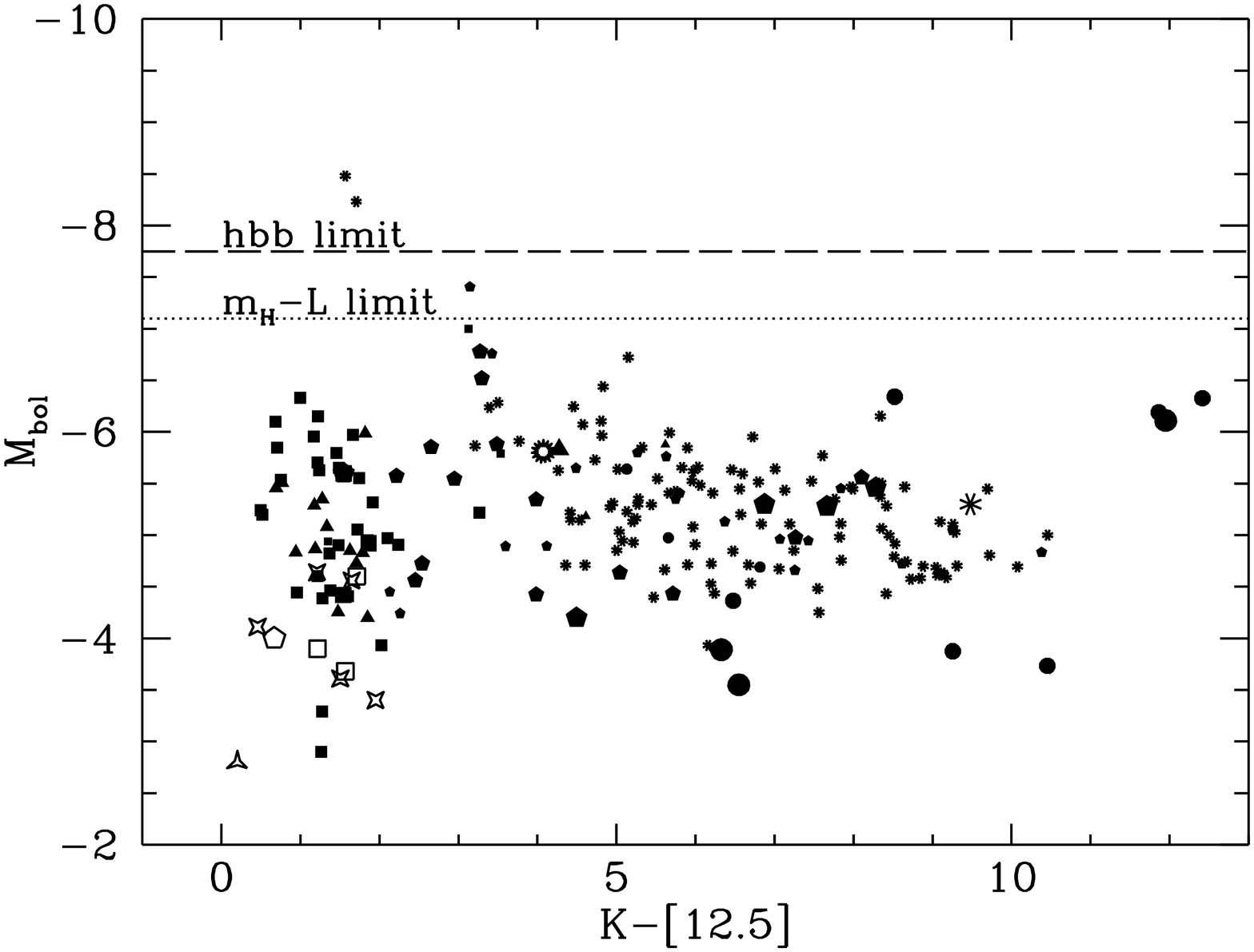}
\end{center}

\end{figure}

\begin{figure}
\begin{center}\includegraphics[width=0.80\columnwidth,angle=-90,clip]{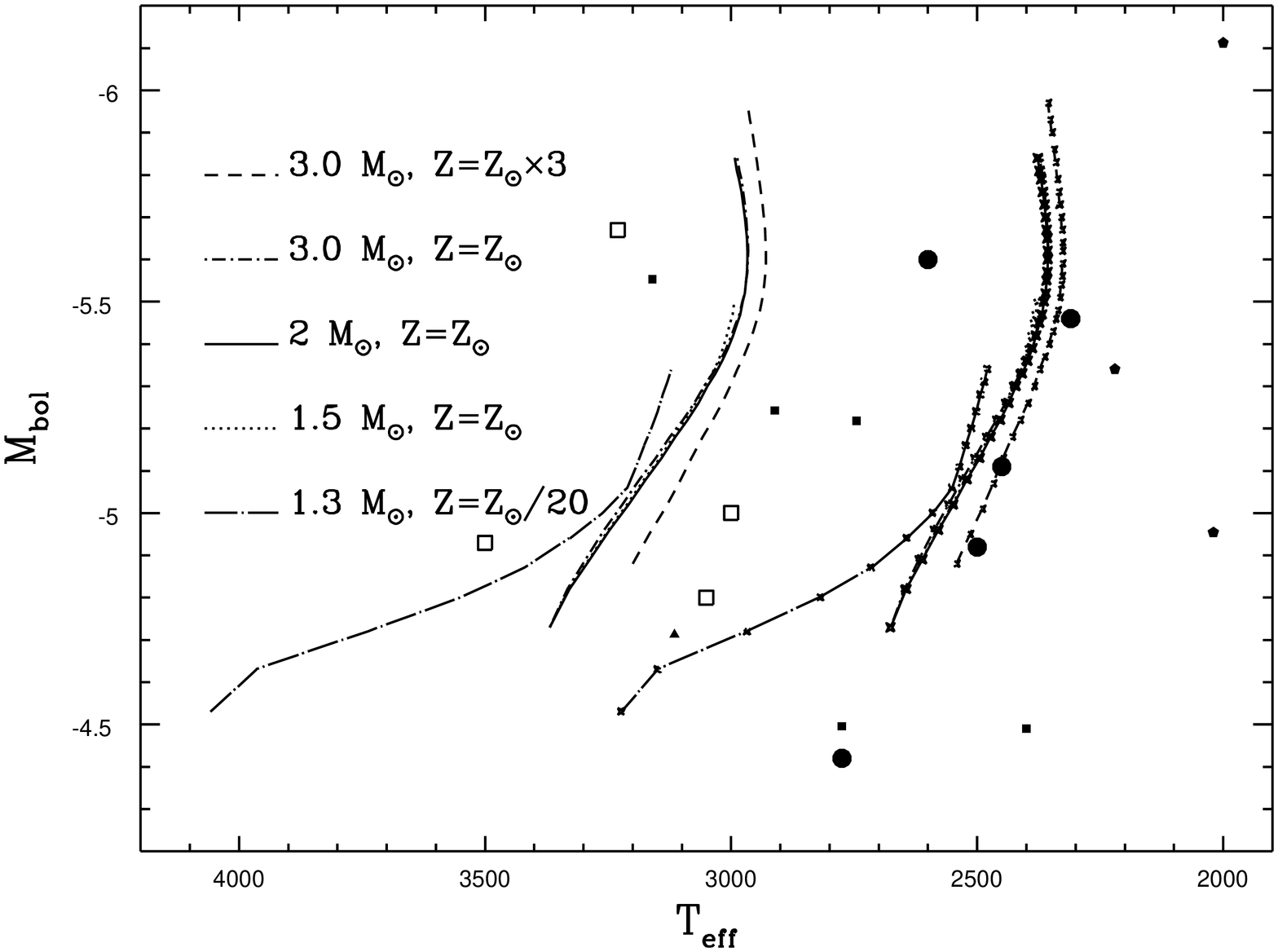}
\end{center}

\end{figure}

\begin{figure*}
\begin{center}\includegraphics[width=18cm,angle=90]{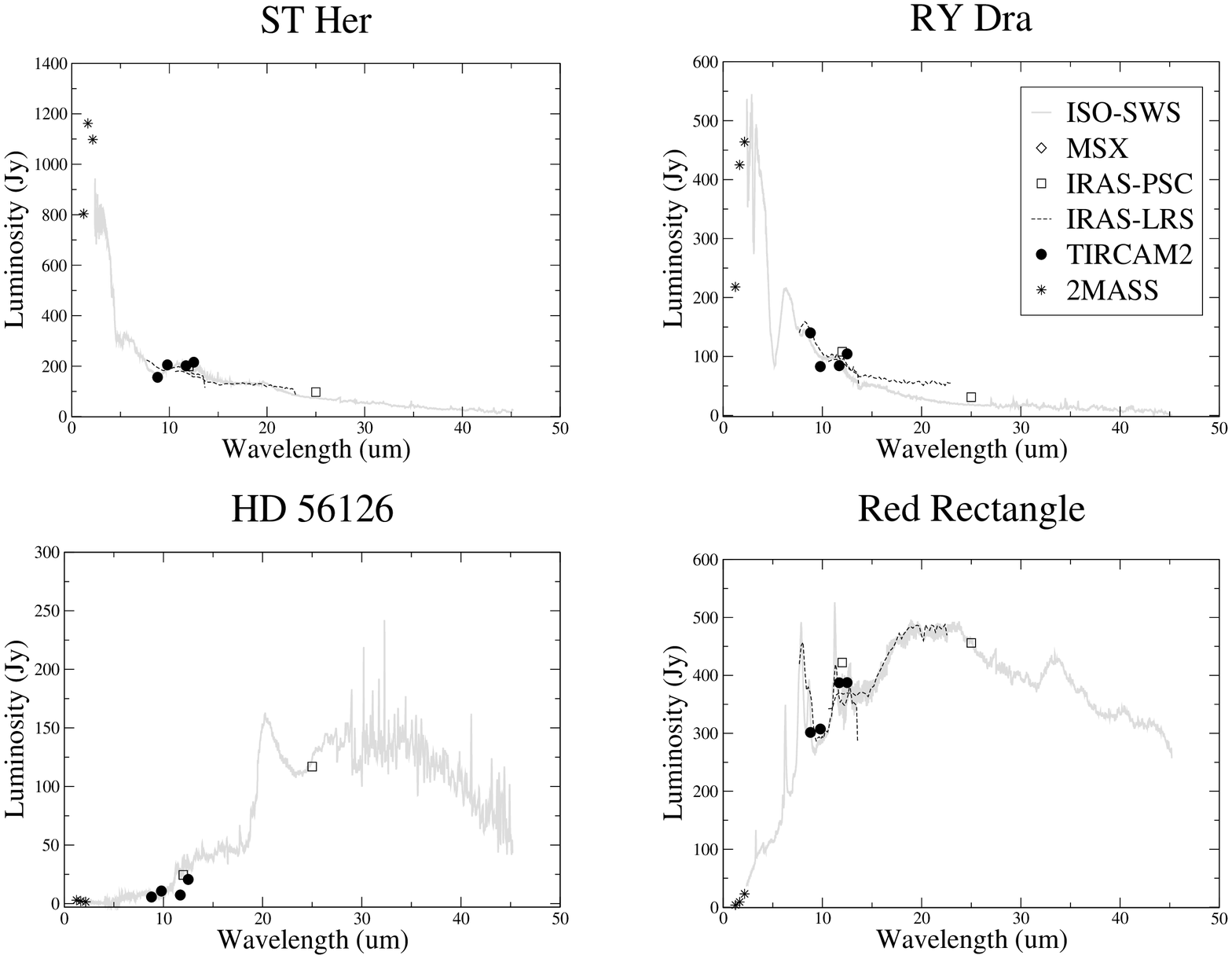}
\end{center}

\end{figure*}

\begin{figure*}
\begin{center}\includegraphics[width=18cm,angle=90]{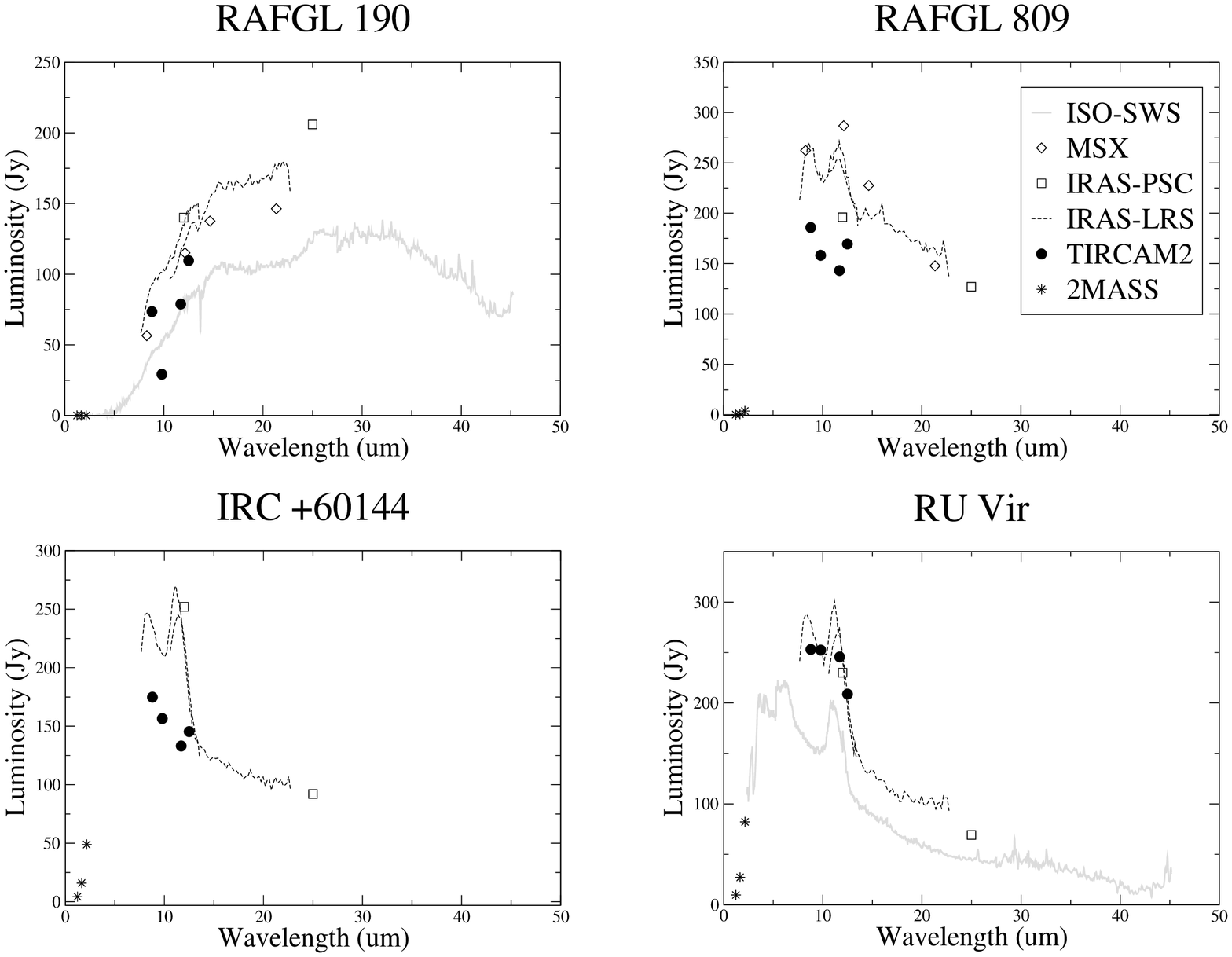}
\end{center}

\end{figure*}

\begin{figure*}
\begin{center}\includegraphics[width=17cm]{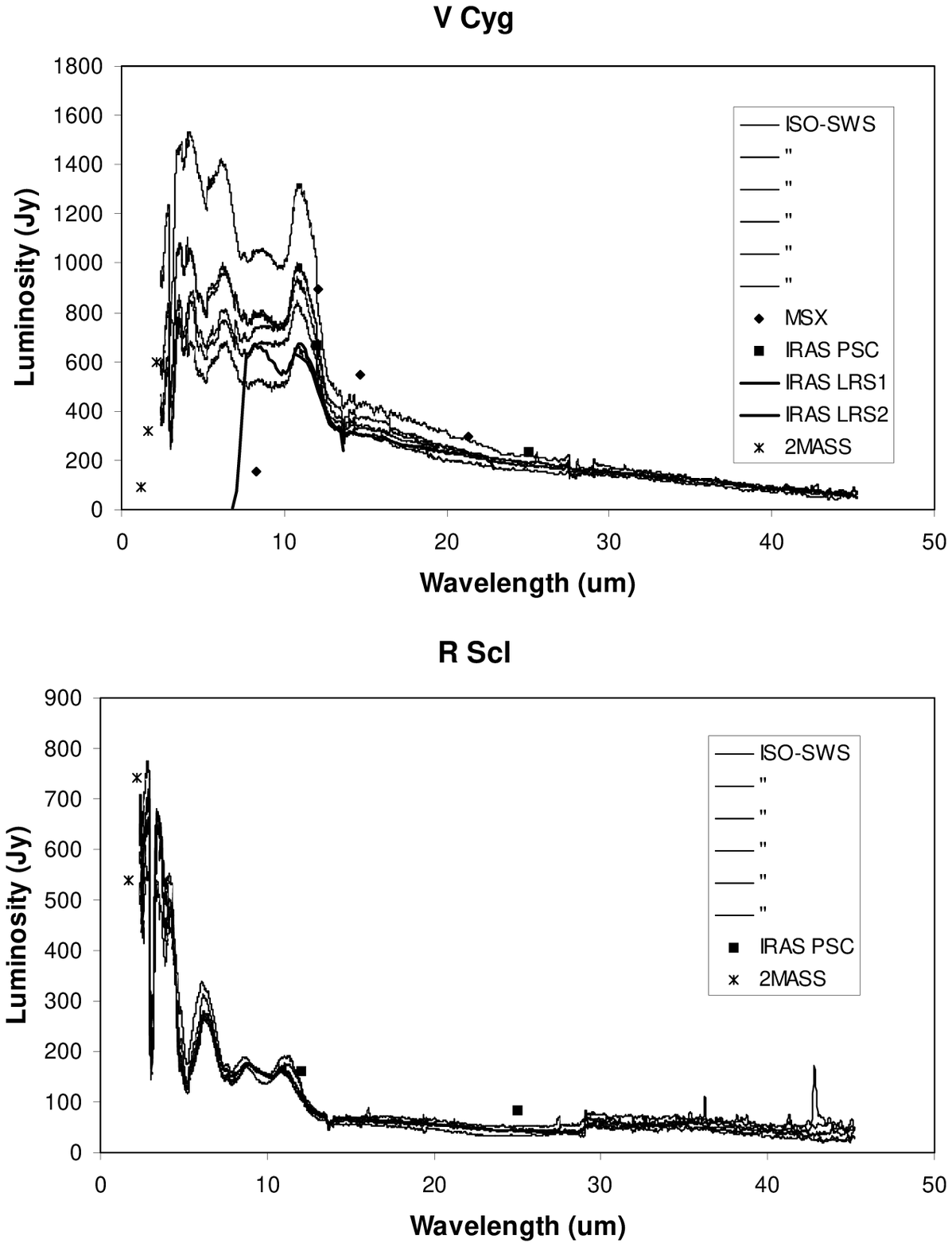}
\end{center}
\end{figure*}

\clearpage \onecolumn

\begin{table}
\caption{The TIRCAM2 Sample of AGB Stars.}
    \small{

         \begin{tabular}{|c|c|c|c|c|c|}
  \hline
  Number & IRAS Name & Other Name & Coordinates & Var. Type & Type \\
  \hline
   1 & 01144+6658  & RAFGL 190 & 01 17 51.62 +67 13 55.4 & P & C \\
  2 & 03186+7016  & RAFGL 482 & 03 23 36.57 +70 27 07.5 & M & C \\
  3 & 04307+6210  & IRC +60144 & 04 35 17.45 +62 16 23.3 & S & C \\
  4 & 04395+3601  & RAFGL 618 & 04 42 53.67 +36 06 53.2 & P & C \\
  5 & 04530+4427  & RAFGL 6319S & 04 56 43.28 +44 32 41.6 & $-$ & C \\
  6 & 05405+3240  & RAFGL 809 & 05 43 49.78 +32 42 06.8 & M & C \\
  7 & 05426+2040  & Y Tau & 05 45 39.41 +20 41 42.1 & S & C \\
  8 & 06012+0726  & RAFGL 865 & 06 03 59.84 +07 25 54.4 & M & C \\
  9 & 06176$-$1036    & Red Rectangle & 06 19 58.22 -10 38 14.7 & P & C \\
 10 & 06291+4319  & RAFGL 954 & 06 32 41.93 +43 17 15.3 & I & C \\
 11 & 07134+1005  & HD 56126 &  07 16 10.26 +09 59 48.0 & P & C \\
 12 & 09452+1330  &   CW Leo  & 09 47 57.38 +13 16 43.7 & M & C \\
 13 & 10131+3049  & CIT 6 & 10 16 02.27 +30 34 18.6 & S & C \\
 14 & 12427+4542  & Y CVn & 12 45 07.83 +45 26 24.9 & S & C \\
 15 & 12447+0425  & RU Vir & 12 47 18.41 +04 08 41.4 & M & C \\
 16 & 12544+6615  & RY Dra & 12 56 25.91 +65 59 39.8 & S & C \\
  \hline
 17 & 06331+1415  & DY Gem & 06 35 57.81 +14 12 46.1 & S & S \\
 18 & 09076+3110  & RS Cnc & 09 10 38.80 +30 57 47.3 & S & S \\
 19 & 12417+6121  & S UMa & 12 43 56.68 +61 05 35.5 & M & S \\
 20 & 15492+4837  & ST Her & 15 50 46.62 +48 28 58.9 & S & S \\
  \hline
 21 & 06297+4045  & IRC +40156 & 06 33 15.75 +40 42 50.9 & P & M \\
 22 & 12277+0441  & BK Vir & 12 30 21.01 +04 24 59.2 & S & M \\
 23 & 13001+0527  & RT Vir & 13 02 37.98 +05 11 08.4 & S & M \\
 24 & 14059+4405  & BY Boo & 14 07 55.76 +43 51 16.0 & I & M \\
 25 & 14219+2555  & RX Boo & 14 24 11.63 +25 42 13.4 & S & M \\
 26 & 14371+3245  & RV Boo & 14 39 15.86 +32 32 22.3 & S & M \\
 27 & 16269+4159  & g Her & 16 28 38.55 +41 52 54.0 & S & M \\
  \hline
\end{tabular}

} \label{tab:one}
   \end{table}

\begin{table}
\caption{Flux Densities of standard stars.}
\par
  \small{
\begin{tabular}{lllll}
\hline
Standard & [8.8]& [9.8] & [11.7] & [12.5]  \\
 & (Jy) & (Jy)& (Jy)& (Jy)  \\
\hline
$\alpha$ Lyr & 49.69 & 40.44  & 28.48 & 25.05 \\
$\beta$ Gem & 152.74 & 120.90 & 88.67 & 76.77 \\
$\alpha$ Boo & 883.72 & 745.32 & 524.94 & 459.01 \\
$\beta$ And & 306.36 & 263.52 & 200.79 & 174.55 \\
$\beta$ Peg & 431.12 & 376.00 & 279.25 & 249.01 \\
$\alpha$ Tau & 752.09 & 646.86 & 481.44 & 419.57 \\
 \hline
\end{tabular}
} \label{tab:two}
\end{table}

\begin{table}
\caption{Our TIRCAM2 observations in the 10$\mu$m
    window.}
    \small{

         \begin{tabular}{|c|c|c|c|c|c|}
  \hline
  Source Number & Epoch & F[8.8] (Jy) & F[9.8] (Jy) & F[11.7] (Jy) & F[12.5] (Jy) \\
  \hline
  1 & 16.1.2003 & 73.5(5.4) & 29.2(8.2) & 78.9(11.3) & 109.6(14.4)  \\
  2 & 16.1.2003 & 147.1(7.9) & $-$ & 146.8(17.6) & 154.6(19.1)  \\
  3 & 16.1.2003 & 174.8(8.8) & 156.5(10.9) & 133.1(16.4) & 145.4(19.2) \\
  4 & 6.12.2003 & 254.6(27.1) & 380.2(42.8) & 490.2(26.7) & 488.0(28.4)  \\
  5 & 13.2.2002 & 87.2(19.1) & 83.0(8.3) & 100.1(10.0) & 87.8(9.0)  \\
  6 & 16.1.2003 & 185.8(9.3) & 158.2(11.9) & 143.1(17.5) & 169.5(21.5)  \\
  7 & 12.2.2002 & 122.6(13.5) & 116.8(7.4) & 133.0(6.7) & 80.9(3.6) \\
  8 & 6.12.2003 & 364.9(28.9) & 414.1(41.1) & 477.9(34.6) & 379.1(43.2) \\
  9 & 7.12.2003 & 301.5(34.2) & 307.5(50.4) & 387.2(35.9) & 387.4(49.0) \\
  10 & 15.1.2003 & 63.4(6.6) &  66.2(14.6) & 129.1(12.6) & 81.6(24.)  \\
  11 & 6.2.2004 & 5.6(1.8) & 10.7(2.4) & 7.3(2.8) & 20.6(2.7)  \\
  12 & 14.1.2001 & 30255.(605.) & $-$ & 41782.(342.) & 38978.(667.9) \\
  13 & 15.1.2003 & 1789.(156.) & 1976.(190.) & 2226.(400.) &  \\
  14 & 6.12.2003 & 342.8(20.3) & 337.0(29.0) & 108.9(22.0) & 220.7(24.6)  \\
  15 & 16.1.2003 & 253.0(27.6) & 252.4(24.3) & 245.7(39.3) & 208.8(25.5)  \\
  16 & 16.1.2003 & 140.1(15.7) & 83.0(10.7) & 84.4(13.8) & 104.4(13.4) \\
  \hline
  17 & 6.2.2004 & 5.3(1.6) & 16.4(3.3) & 34.7(10.5) & 24.5(4.5) \\
  18 & 14.1.2001 & 398.4(40.0) & 550.0(82.5) & 453.1(63.4) & 378.3(57.0) \\
   "  & 11.2.2002 & 568.5(47.8) & 790.7(29.8) & 497.2(13.0) & 452.1(33.0)  \\
   "  & 6.12.2003 & 570.0(30.0) & 739.1(49.0) & 530.0(37.0) & 477.1(28.0)  \\
  19 & 11.2.2002 & 4.5(0.5) & $-$ & 2.8(0.5) & 3.1(0.7)  \\
  20 & 4.2.2004 & 155.7(8.0) & 205.2(33.0) & 201.4(50.0) & 215.5(28.0) \\
  \hline
  21 & 15.1.2003 & 209.4(10.8) & 334.1(37.8) & 321.4(69.0) & 134.9(10.0)  \\
  22 & 4.2.2004 & 224.6(11.3) & 217.6(32.6) & 118.4(29.6) & 211.6(27.5) \\
  23 & 14.1.2001 & 365.7(16.4) & 496.4(24.0) & 394.3(22.4) & 432.5(54.3) \\
  24 & 12.2.2002 & 77.7(8.6) & 72.2(1.3) & 51.6(2.6) & 40.2(2.3)  \\
  25 & 11.2.2002 & 694.8(58.4) & 922.2(34.7) &  795.8(20.5) & 729.2(52.9)  \\
  26 & 6.2.2004 & 106.7(13.9) & 151.3(47.2) & 75.1(21.0) & 86.4(8.6)  \\
  27 & 12.2.2002 & 461.6(50.9) & 427.4(27.0) & 340.4(17.3) & 293.6(13.2)  \\
  \hline
\end{tabular}

}
         \label{tab:three}
   \end{table}

\begin{table}
\caption{Source fluxes in the 2MASS, MSX \& ISO filters (Jy)}
         \begin{tabular}{|c|c|c|c|c|c|c|c|}
  \hline
Source  no.  &   Var. type    &   J   &   H   &   K   &   14,6    &   21,3    &   Data orig.   \\
\hline
 1 &   P   &   0.01    &   0.02    &   0.01    &   96.3    &   110 &   ISO \\
 2 &   M   &   0.09    &   1.1 &   6.8 &   $-$ &   $-$ &   $-$ \\
 3 &   S   &   4.3 &   16.0    &   48.9    &   $-$ &   $-$ &   $-$ \\
 4 &   P   &   $<$ 0.01    &   0.03    &   0.20    &   651 &   1260    &   ISO \\
 5 &   $-$ &   $<$ 0.01    &   0.03    &   0.35    &   62.0    &   51.2    &   MSX \\
 6  &   M   &   0.05    &   0.53    &   3.6 &   227.5   &   147.8   &   MSX \\
 7 &   S   &   242 &   495 &   481 &   59.1    &   36.4    &   MSX \\
 8  &   M   &   $<$ 0.01    &   0.20    &   2.5 &   $-$ &   $-$ &   $-$ \\
 9 &   P   &   3.7 &   9.0 &   23.0    &   377 &   472 &   ISO \\
10 &   I   &   8.1 &   22.7    &   39.3    &   $-$ &   $-$ &   $-$ \\
11  &   P   &   2.9 &   2.1 &   1.5 &   42.0    &   116 &   ISO \\
12 &   M   &   2.7 &   74.9    &   469 &   26309   &   18194   &   ISO \\
13 &   S   &   2.5 &   17.1    &   91.4    &   $-$ &   $-$ &   $-$ \\
14 &   S   &   641 &   1331    &   1316    &   53.8    &   33.5    &   ISO \\
15 &   M   &   9.7 &   27.2    &   82.2    &   91.6    &   54.3    &   ISO \\
16 &   S   &   218 &   425 &   464 &   50.4    &   24.7    &   ISO \\
    \hline
17 &   S   &   85.4    &   156 &   143 &   12.7    &   7.8 &   MSX \\
18 &   S   &   3065    &   4324    &   3743    &   $-$ &   $-$ &   $-$ \\
19 &   M   &   26.3    &   43.4    &   41.4    &   $-$ &   $-$ &   $-$ \\
20 &   S   &   804 &   1162    &   1098    &   149 &   104 &   ISO \\
    \hline
21 &   P   &   11.5    &   33.8    &   78.3    &   $-$ &   $-$ &   $-$ \\
22 &   S   &   966 &   1448    &   1306    &   $-$ &   $-$ &   $-$ \\
23 &   S   &   1145    &   1843    &   1770    &   $-$ &   167 &   ISO \\
24 &   I   &   630 &   1033    &   826 &   $-$ &   $-$ &   $-$ \\
25 &   S   &   $-$ &   4210    &   3948    &   584 &   422 &   ISO \\
26 &   S   &   512 &   785 &   700 &   76  &   66  &   ISO \\
27 &   S   &   3841    &   5627    &   4760    &   273 &   150 &   ISO \\
\hline
\end{tabular}

\label{tab:four}
   \end{table}

\begin{table}
\caption{Relevant parameters and distances of the sample stars.}

\begin{tabular}{|c|c|c|c|c|c|c|}
 \hline
Source no.  &   Var. type   &  $\dot M$ [$M_{\odot}/yr$]   &   Type (ref.)   &   v [km/s]  &   d [kpc]  &   Type ref. \\
\hline
 1  &   P   &   2.45E-05    &   L   &   18.2    &   2.79    &   G  \\
 2  &   M   &   1.05E-05    &   L   &   12.2    &   1.97    &   G  \\
 3 &   S   &   6.33E-06    &   L   &   18.5    &   1.03    &   G  \\
 4  &   P   &   2.00E-04    &   Mei1998 &   19.5    &   1.70    &   Mei1998 \\
 5  &   $-$ &   1.46E-05    &   L   &   20.2    &   2.60    &   G  \\
 6  &   M   &   2.40E-05    &   L   &   28.0    &   2.01    &   G  \\
 7 &   S   &   1.60E-06    &   B   &   11.0    &   0.74    &   B  \\
 8  &   M   &   1.16E-05    &   L   &   15.8    &   1.47    &   G  \\
 9  &   P   &   1.00E-04    &   Men2002 &   5.0 &   0.71    &   Men2002 \\
10  &   I   &   6.37E-06    &   L   &   21.4    &   2.19    &   G  \\
11  &   P   &   2.23E-05    &   L   &   10.7    &   2.40    &   Hon2003 \\
12 &   M   &   3.30E-05    &   B   &   14.7    &   0.15    &   B  \\
13 &   S   &   6.50E-06    &   B   &   17.0    &   0.41    &   B  \\
14 &   S   &   1.40E-07    &   B   &   8.5 &   0.26    &   B  \\
15 &   M   &   2.30E-06    &   B   &   18.4    &   0.68    &   B  \\
16 &   S   &   4.40E-07    &   B   &   10.0    &   0.55    &   B  \\
\hline
17 &   S   &   6.27E-08    &   G-dJ    &   8.0 &   0.56    &   G-dJ  \\
18 &   S   &   3.92$-$5.20E-08 &   G-dJ    &   7.2 &   0.12    &   Hip \\
19 &   M   &   $-$ &   $-$ &   $-$ &   1.15    &   G-dJ  \\
20 &   S   &   4.52$-$7.21E-08 &   G-dJ    &   9.1 &   0.31    &   Hip \\
\hline
21  &   P   &   1.00E-05    &   L   &   16.3    &   1.60    &   L  \\
22 &   S   &   5.49E-07    &   W   &   7.5 &   0.18    &   Hip  \\
23 &   S   &   3.70E-07    &   L   &   9.3 &   0.14    &   Hip  \\
24 &   I   &   $-$ &   $-$ &   $-$ &   0.14    &   Hip  \\
25 &   S   &   6.48E-07    &   L   &   10.2    &   0.16    &   Hip  \\
26 &   S   &   5.49E-07    &   L   &   8.1 &   0.39    &   Hip  \\
27  &   S   &   1.43E-07    &   L   &   8.5 &   0.11    &   Hip  \\
\hline
\end{tabular}

\tablecomments{Here the following rules apply for labelling. i)
For the variability type, M means Mira; S means Semiregular; I
means Irregular; P is used for all Post-AGB stars; ''-'' for all
stars of unknown variability. ii) References for mass loss are
quoted in the  following synthetic way: B \citep{berg05}; L
\citep{loup}; G-dJ \citep{groenewegen98b}; Mei1998
\citep{meixner}; Men2002 \citep{menshchikov}; W \citep{winters03}.
iii) Synthetic references for distances have the following
assignment: Hip (Hipparcos observations); B \citep{berg05}; L
\citep{loup}; G \citep{groenewegen02}; G-dJ
\citep{groenewegen98b}; Mei1998 \citep{meixner}; Men2002
\citep{menshchikov}; Hon2003 \citep{hony}.} \label{tab:five}
   \end{table}

\end{document}